\documentclass[aps,prb,twocolumn,floatfix]{revtex4}
\usepackage{color,graphicx,pstricks}
\begin{document}

\title {Magnetic Reconstructions in B-site Doped Manganites}

\author{Kalpataru Pradhan}

\affiliation{CMP Division, Saha Institute of Nuclear Physics, 
1/AF Bidhan Nagar, Kolkata 700064, India}


\begin{abstract}
The magnetic nature of the B-site dopants controls the magnetic phases in 
B-site doped manganites RE$_{1-x}$AE$_{x}$Mn$_{1-\eta}$B$_{\eta}$O$_3$.
Different B-site dopants of equal valence, doped into the same reference manganite,
lead to different magnetic phases at low temperature, which can not be explained
using the valence change scenario. We focus on trivalent B-site dopants in
CE-CO-OO-I manganites at half-filling $x=0.50$ to study the role of magnetic
interactions between the B-site dopants and the neighboring Mn-sites by using
a two-orbital double-exchange model including super-exchange
interactions, Jahn-Teller lattice distortions and substitutional disorder
in two dimensions. We show that the magnetic reconstructions around the B-site
dopants due to the modified double-exchange and super-exchange interactions
control the phase competition in B-site doped manganites.
\end{abstract}

\maketitle

In manganites~\cite{Dagotto-2005,Tokura-2000,Tokura-2006}, 
RE$_{1-x}$AE$_x$MnO$_3$ (RE and AE denote rare-earth and alkaline-earth elements), 
A-site disorder~\cite{Attfield-1998,Attfield-2001} is unavoidable 
except in few cases of specific doping $x$ and special growth 
technique~\cite{Mathieu-2004,Akahoshi-2003}. A colossal response in a magnetic 
field emerges due to the A-site disorder which promotes phase 
coexistence and metal-insulator transition~\cite{Moreo-2000,Kumar-2004}. Similar 
phase coexistence scenario have also been observed in B-site doped manganites 
RE$_{1-x}$AE$_{x}$Mn$_{1-\eta}$B$_{\eta}$O$_3$ where a few percentage of Mn-sites 
are replaced by foreign elements named B-site dopants
~\cite{Kimura-1999,Kimura-2000,Moritomo-1999,Mori-2003,Moritomo-2004,Machida-2002,
Ahn-1997,Barnabe-1997, Sakai-2008,Dhiman-2013,Lu-2014}. In addition, the B-site disorder can tune 
a ferromagnetic (FM) metal to an anti-ferromagnetic (AF) 
insulator~\cite{Ahn-1997,Sakai-2008} or vice-versa~\cite{Kimura-2000}. 
This promotion of a competing ordered phase has no equivalent in the case of A-site 
disorder.

In few cases the B-site dopant driven transition from a FM-M to a charge ordered 
insulator or the reverse transition can be explained using the valence change 
scenario~\cite{Pradhan-2008}. Here the Mn valence changes to 
$3 + \nu$ for an $\alpha$-valent B-dopant in the reference manganites
RE$^{3+}_{1-x}$AE$^{2+}_{x}$Mn$^{3+\nu}_{1-\eta}$B$^{\alpha}_{\eta}$O$^{2-}_3$,
where the charge neutrality requirement, 
$\nu(\eta, \alpha, x) = (x + \eta(3 -\alpha))/(1 -\eta)$. The new effective 
hole density $x_{eff} = \nu$. 
Significantly, all B-site doped experiments {\it can not} be explained by the 
valence change argument as some B-site dopants, particularly the magnetic ones 
with same valency behave differently~\cite{Moritomo-1999,Kimura-2000,Hebert-2002}. 
This requires a careful analysis of magnetic interactions between the B-site 
dopants and the neighboring Mn-sites. 

\begin{figure}[!t]
\centerline{
\includegraphics[width=8.7cm,height=5.8cm,clip=true]{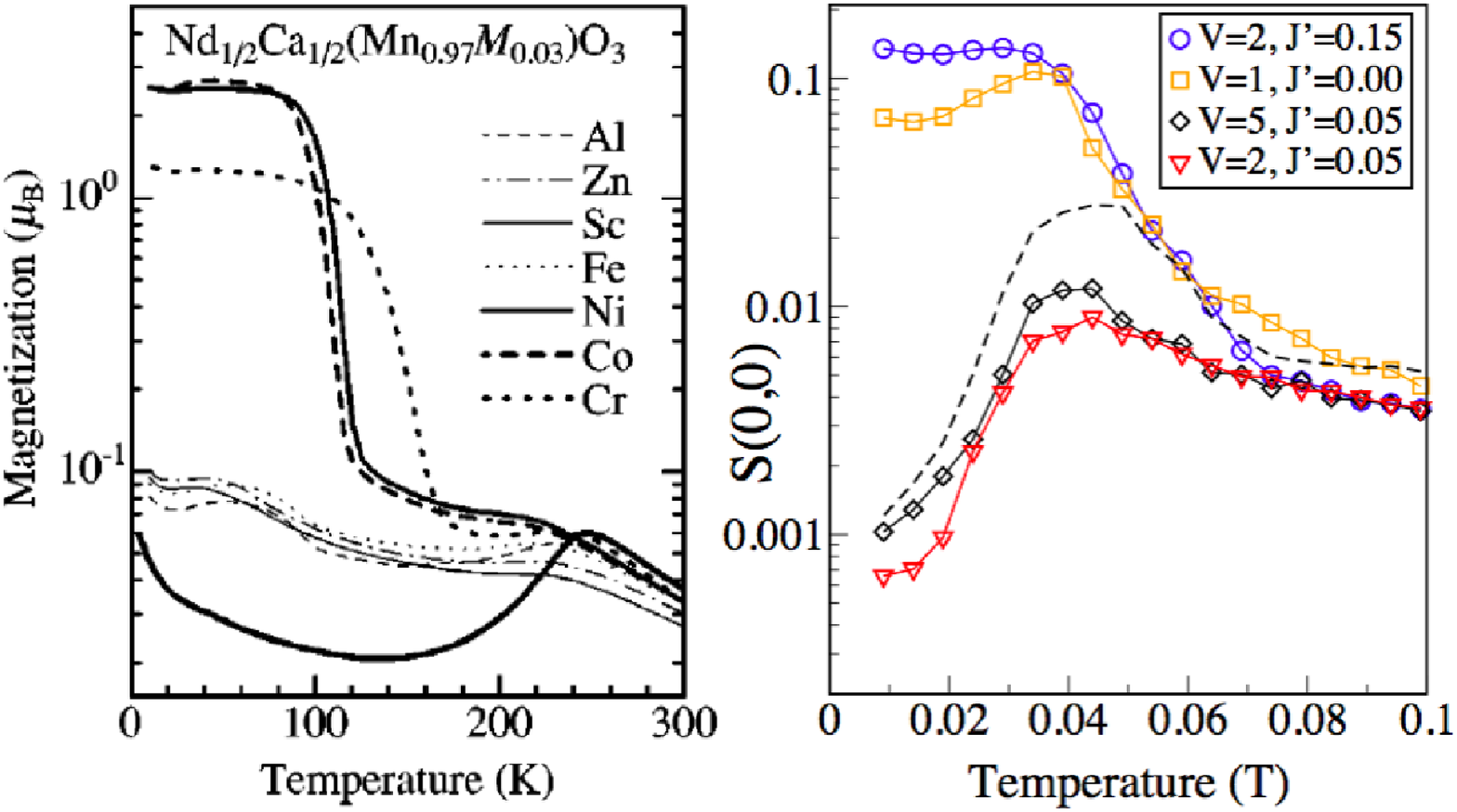}}
\caption{
Left: Temperature dependence of magnetization for different 
B-site dopants in  Nd$_{0.5}$Ca$_{0.5}$MnO$_3$. Magnetization is measured 
under a field of 0.5 T [results from  A. Machida, 
{\it et al}., Phy. Rev. B {\bf 65}, 064435 (2002)].
Right: Temperature dependence of the FM
peak [$S({\bf 0,0})$] for four representative parameter points 
in the phase diagram (Fig-\ref{pd_J_V.eps}) with $h = 0.002$ 
(lattice size: $24 \times 24$). The dotted line is 
for the reference CE-CO-OO-I phase.}
\label{vj_fm_expt.eps}
\end{figure}

Our focus is mainly on $3+$ dopants in a CE-CO-OO-I (CO:charge order; OO:Orbital 
order; I: Insulator) manganite at $x=0.50$. 
Some elements from our $3+$ dopants may exist in $2+$ state due to the presence 
of mixed valence, but that will not affect the qualitative results of this 
paper. Let us briefly discuss the available experimental data that emphasize 
the magnetic character of the B-site dopants. The FM-M 
ground state in Cr doped La$_{0.5}$Ca$_{0.5}$MnO$_{3}$ (LCMO)~\cite{Moritomo-1999} 
cannot be explained using the valence change argument. With Cr, a $3+$ dopant, 
the hole density of the manganite {\it increases} from the initial 
$x = 0.50$ but there is no FM-M phase which has hole density greater than 
$x=0.50$~[\citenum{Tokura-2000}] in LCMO.  When Nd$_{0.5}$Ca$_{0.5}$MnO$_{3}$ (NCMO), 
another CE-CO-OO-I with smaller bandwidth than LCMO, doped with Cr, shows 
coexistence of FM-M and CO-I phases at low 
temperature~\cite{Kimura-2000, Machida-2002}. There is no FM phase for $x > 0.5$ 
in NCMO either. Both these parent manganites 
have FM-M phases at $x < 0.50$. Dark field images at temperatures below the 
transition temperature shows that the hole density in the FM domains is less 
than $x=0.50$~[\citenum{Moritomo-1999}]. This is surprising because the valence 
change on doping $3+$ dopants is in a direction opposite to the clean 
ferromagnetic (FM) phase, so Cr doping should not have led to a FM-M. 
In addition to Cr, Ni and Co also lead to ferromagnetism at low 
temperature when doped into NCMO, going against the valence change 
scenario. Note, however,  that $3+$ dopants like Fe, Al, and Sc 
when doped into the same manganite at $x=0.50$ do not induce ferromagnetism 
at any temperature as shown in the left hand side of the 
Fig-\ref{vj_fm_expt.eps}~ [from ref.\citenum{Machida-2002}]. Cr, Ni, and Co also 
induces ferromagnetism at low temperature at the expense of AF ground state in 
Pr$_{0.5}$Ca$_{0.5}$MnO$_{3}$ (PCMO)~\cite{Hebert-2002}.

In this letter, we analyze the importance of magnetic interactions between the 
B-site dopants and the neighboring Mn-sites to understand different magnetic 
phases at the low temperature. The magnetic reconstructions around the B-site 
dopants control the phase competition between the FM and the AF phase. We 
explain the origin of two type of magnetic reconstructions that induces 
ferromagnetism in an AF insulator. We use a two-dimensional model Hamiltonian 
for manganites with B-site dopants in the limit 
$J_H \rightarrow \infty$~[\citenum{Dagotto-2001}]. The model is given by 

\begin{eqnarray}
H_{tot}&=& H_{ref} + H_{imp}, \hspace{0.1cm} \textrm{where} \cr 
H_{ref}&=&\sum_{\langle ij \rangle \sigma}^{\alpha \beta}
{\tilde t}_{\alpha \beta}^{ij}
 d^{\dagger}_{i \alpha \sigma} d^{~}_{j \beta \sigma}
+ J\sum_{\langle ij \rangle} {\bf S}_i.{\bf S}_j \cr
&&  ~~- \lambda \sum_i {\bf Q}_i.{\mbox {\boldmath $\tau$}}_i
+ {K \over 2} \sum_i {\bf Q}_i^2  \hspace{0.2cm} \textrm{and} \cr  
H_{imp}&=&V \sum_{n \alpha \sigma} 
d^{\dagger}_{n \alpha \sigma}d^{~}_{n \alpha \sigma} 
+ J'\sum_{\langle n j \rangle} {\bf S}_{n}.{\bf S}_j 
+ V_c \sum_{\langle n j \rangle} q_n q_j. \cr
\nonumber
\end{eqnarray}

The reference `manganite model' $H_{ref}$ involves the nearest neighbor 
hopping of $e_g$ electrons with amplitude ${\tilde t}^{ij}_{\alpha \beta}$ 
(two orbitals a and b), anti-ferromagnetic (AF) super-exchange (SE) $J$ between Mn 
t$_{2g}$ spins, and Jahn-Teller (JT) interaction $\lambda$ between 
the electrons and the phonon modes ${\bf Q}_i$ in the adiabatic limit. 
The hopping amplitudes ${\tilde t}^{ij}_{\alpha \beta}$ depend upon the 
orientations of $t_{2g}$ spins at sites $i$ and $j$ where
${\tilde t}^{ij}_{\alpha \beta}$ = $\Theta_{ij}t^{ij}_{\alpha \beta}$
with
  $\Theta_{ij} \!=\! \cos (\theta_{i}/2)\cos (\theta_{j}/2)                             
  \!+\! \sin (\theta_{i}/2)\sin (\theta_{j}/2)                                                   
  e^{-i(\phi_{i}-\phi_{j})}.$ 
We treat all t$_{2g}$ spins and phonon degrees of freedom as 
classical~\cite{Dagotto-1998}, and  measure all energies in units 
$t_{aa}$=1. We set stiffness of the JT modes, $K$=1 and $|{\bf S}_i|=1$. 
The overall carrier density is controlled through the chemical potential. 
For details please see Ref.\citenum{Dagotto-2001, Pradhan-2007, 
Pradhan-2008}
 
We use an exact diagonalization method to the e$_g$ electrons in the 
presence of the t$_{2g}$ spins and the phonon modes (Q). A Monte Carlo 
(MC) technique based on the `travelling cluster approximation' (TCA) 
\cite{1-Kumar-2006} is employed to access large system 
sizes~\cite{Pradhan-2007,3-Kumar-2006,Kumar-2008,Pradhan-2013}. 
$H_{ref}$ reproduce the correct sequence of magnetic phases in the `clean' 
manganites for $J=0.1$ and $\lambda=1.6$~[\citenum{Pradhan-2008}].

\begin{figure}[!t]
\centerline{ 
\includegraphics[width=8.5cm,height=6.0cm,clip=true]{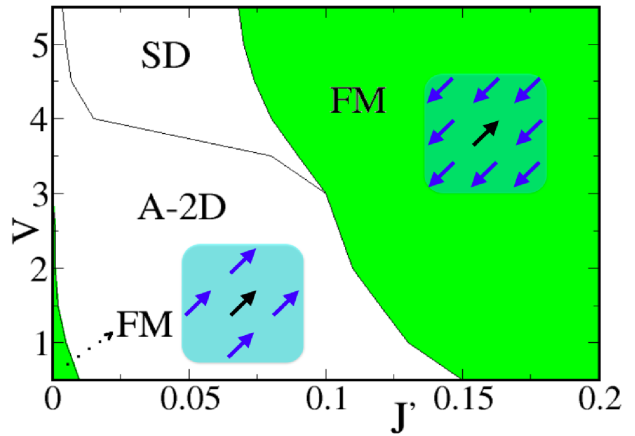}}
\caption{Low temperature (T = 0.01) `phases' at $x=0.50$ for varying $V$ and
$J'$ for $3+$ B-site dopants in a external magnetic field $h=0.002$. We have 
taken $\lambda=1.6$ and $J=0.1$. There are two regions in the phase diagram 
where ferromagnetism is induced. The other regions are either spin disordered 
phase or mostly A-2D type phase.}
\label{pd_J_V.eps}
\end{figure}

The CE-CO-OO-I phase is stable only at $x=0.50$ for $\lambda=1.6$ and 
$J=0.1$~[\citenum{Yunoki-2000,Pradhan-2007}]. There are phase 
separation windows (as shown in Fig-2 of Ref. \citenum{Pradhan-2008}) on both 
sides of the $x=0.50$ CE-CO-OO-I phase e.g., anti-ferromagnetic A-type phase 
(A-2D) phase for $x \ge 0.55$ and FM-M phase for $x \le 0.40$. A few percentage 
of $3+$ dopants at $x=0.50$ will shift the $x_{eff}$ into the phase separation 
region between $x=0.50$ and $x=0.55$ creating a mixture of A-2D and CE-type phase 
without any ferromagnetic correlations. The valence change argument may be 
the correct way to explain the very weak ferromagnetic feature when Fe, Al or 
Ga dopants are introduced into NCMO (see left side of Fig-\ref{vj_fm_expt.eps}).
But, the mystery is why some trivalent/divalent dopants like Cr and Ni 
induce ferromagnetism in the CE-CO-OO-I phase, while other $3+$ dopants such as 
Fe unable to do so.

We modifiy the local physics around the B-site dopant. In principle the SE 
interaction between the B-Mn sites can be very different from the SE interaction 
between the Mn-Mn sites. The position of the impurity level at the 
B-site dopant also makes a difference since it controls double-exchange (DE) driven 
ferromagnetic coupling between the B-Mn sites. Another important aspect is short-range 
Coulomb interaction between the B-site dopant and the neighboring Mn site, particularly 
when charge ordered reference states are considered. When a B-site dopant is introduced 
into a CE-CO-OO-I state the fixed charge state of the B-site dopant forces a rearrangement 
in the valence of the neighboring Mn to minimize the Coulomb repulsion~\cite{Fratini-2001}.
For instance, a $3+$ dopant like Cr is more likely to be surrounded by Mn$^{4+}$ ions. 
$H_{imp}$ contains these three changes in the model due to the B-site dopants; 
i) the B-site dopant e$_g$ states placed at an energy V above the center 
of the Mn band, ii) the SE coupling is modified to J' between the B-site moments S$_n$ and 
the neighboring Mn moments, and iii) a nearest neighbor (NN) Coulomb repulsion V$_c$ between 
the B-site dopant and the neighboring Mn sites is added~[\citenum{Pradhan-2008}]. We assume 
$V_c = 0.1$ in our calculation~\cite{Fratini-2001,Shenoy-2007}. The quantitative physics 
does not change without the NN Coulomb repulsion ($V_c = 0$). 

The two parameters $V$ and $J'$ define our minimal set to tune the effective magnetic 
interaction between the B-site dopants and neighboring Mn-sites. Fig-\ref{pd_J_V.eps} 
shows the phase diagram of the B-site doped CE-CO-OO-I 
for different combinations of $V$ and $J'$. We set $\eta = 0.08$, and 
have used a small external magnetic field $h=0.002$ as in the experiments 
(see Fig-\ref{vj_fm_expt.eps}). In an external 
magnetic field a Zeeman coupling $H_{mag} = - h\cdot\sum_i {\bf S}_i$ 
is added to the Hamiltonian. We find two unexpected FM regions, 
one at lower $J'$ and $V$ values, the other at higher values of $J'$.
The non-ferromagnetic region is divided into two parts: i) A-2D type correlations 
and ii) spin disordered (SD) like phase. 

We believe that the ferromagnetism at large SE regime is due to the 
complex structure of CE-CO-OO-I phase. The Mn$^{3+}$ (or Mn$^{4+}$) ion in 
CE-CO-OO-I phase (checker-board arranged Mn$^{3+}$ and Mn$^{4+}$ ions) has 
two NN spins aligned parallel to it and other two NN spins aligned 
anti-parallel to it. For Mn$^{3+}$, out of four next NN spins, two are 
parallel and other two are anti-parallel to the central Mn spin. 
Due to large J' at B-site, 
all the neighboring Mn spins align anti-parallel to the B-site moment. To gain 
kinetic energy along a `1+8' ring all four next NN align anti-parallel 
to the B-site  as shown in the inset of Fig-\ref{pd_J_V.eps}. For Mn$^{4+}$, 
all four next NN are already aligned anti-parallel to it. At large 
enough impurity concentration the FM `1+8' rings align in the same direction 
to promote FM correlations in a AF phase. Notice, this argument 
does not involve the valence of the dopant.

\begin{figure}[!t]
\centerline{
\includegraphics[width=8.5cm,height=8.30cm,clip=true]{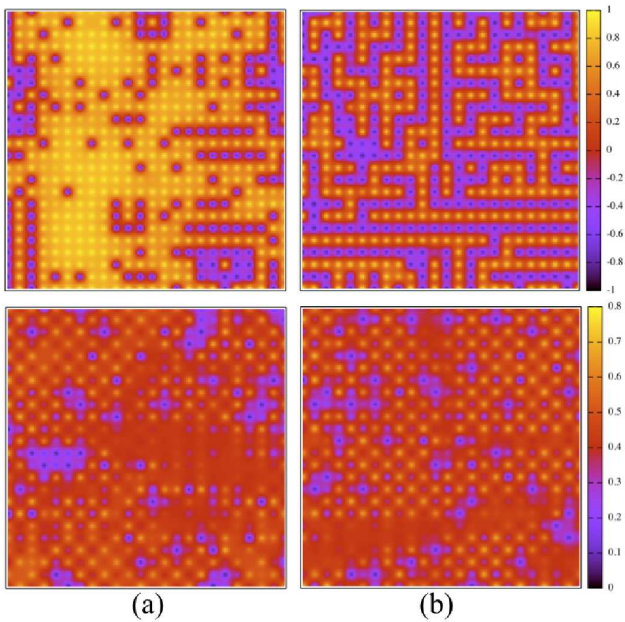}}
\caption{
The $z$ components of the t$_{2g}$ spins (top row) and the electron density 
(bottom row) for each site on a $24 \times 24$ lattice at T = 0.01 ($\eta=0.08$).
(a) $V=2$ and $J'=0.15$, (b) $V=5$ and $J'=0.05$.
} 
\label{mag_cr_fe.eps}
\end{figure}

Along with the AF coupling $J'$, the other crucial effect comes 
from the short range Coulomb repulsion $V_{c}$. 
Due to this $3+$ B-site dopants prefer Mn$^{4+}$ ions as nearest 
neighbors. The effective hole density increases with $3+$ 
dopant but most of the low electron density (high hole density) 
sites are close to the B-site dopants. Some sites with high electron 
density (low hole density) club together in impurity free patches. 
The overall pattern that emerges is a complex mixture of FM-M and AF 
regions. Fig-\ref{mag_cr_fe.eps} (a) shows the $z$ components of t$_{2g}$ 
spins and the electron density for each site from MC snapshots for 
$\eta = 0.08$. The MC snapshot shows ferromagnetic 
patches and we found that the effective hole density of these 
ferromagnetic patches are less than 0.50, but are weakly charge ordered, 
unlike in the experiment~[\citenum{Moritomo-1999}]. This ferromagnetism is 
purely due to the magnetic and the electronic reconstructions around the 
B-site dopants. We believe that the magnetic reconstructions also lead to 
ferromagnetism in Cr doped off-half-filled manganites 
{\it e.g}, 
(La$_{0.3}$Pr$_{0.7}$)$_{0.65}$Ca$_{0.35}$Mn$_{1-\eta}$Cr$_{\eta}$O$_3$
~[\citenum{Dhiman-2013}], 
La$_{0.4}$Ca$_{0.6}$Mn$_{1-\eta}$Cr$_{\eta}$O$_3$~[\citenum{Lu-2014}].

For lower $V$ and $J'$ values there is another narrow ferromagnetic 
window. For lower $J'$, B-site dopant connects to nearest neighbors 
Mn-sites ferromagnetically in the absence of significant SE interaction. 
This FM alignment is due to the gain in DE energy for smaller V values. 
In the present case, clusters of `5 sites' spins forms a large 
magnetic moment, as shown in the inset of Fig-\ref{pd_J_V.eps}. 
This `5 sites' spin structure is the 
building block for ferromagnetism in this regime. In fact, a ferromagnetic 
SE interaction will broaden this ferromagnetic region. As 
the impurity level ($V$) grows the FM phase gets quickly suppressed due 
to the reduction in the DE energy gain. 

In the phase diagram there is no ferromagnetic phase at higher $V$ and lower 
$J'$ values. In this case the system cannot generate either a cluster of 
`5 sites', or the `$1+8$' ferromagnetic configuration. The spin disordered (SD) 
region is a complex mixture of A-2D phase and the CE phase. 
Fig-\ref{mag_cr_fe.eps} (b) shows the $z$ components of t$_{2g}$ spins 
and electron density for each site for $V=5$ and $J'=0.05$. There is no 
significant ferromagnetic correlations. Other dominant phase (for lower $V$ and 
intermediate $J'$ values) is A-2D type phase. This phase is expected due to 
the valence change scenario discussed earlier. 

It clear from the experiments~\cite{Machida-2002,Hebert-2002} that doping 
Al, Fe, Ga, and Sc (say type-I) in CE-CO-OO-I manganite do not induce any 
ferromagnetism, whereas Cr, Ni, and Co (say type-II) lead to 
ferromagnetism at low temperature at the expense of AF ground state. 
Type-I dopants either have $d^0$ or 
$d^{10}$ configuration except Fe which has $d^5$ electronic 
configuration while $d$ orbitals are partially occupied in type-II dopants. 
Based on the electronic configuration $3+$ dopants can be divided into 
magnetic and non-magnetic. Magnetic dopants interact magnetically with 
their neighboring Mn-sites. Experimental result suggests 
that Cr interacts antiferromagnetically to the Mn-sites whereas Ni couples
ferromagnetically~\cite{Studer-1999}. Fe is magnetic but interacts weakly 
with the Mn-sites due to its stable $d^{5}$ configuration~\cite{SE-U} and 
this may be the cause for the absence of ferromagnetism on Fe doping.
Other dopants like Al, Sc, and Ga which do not have partially filled 
$d$ electrons are categorized as non-magnetic. These non-magnetic 
dopants do not have any magnetic interaction (valid for large $V$ and $J'=0$) 
with Mn-sites. The physics due to the valence change scenario dominates 
for these non-magnetic $3+$ dopants and there is no ferromagnetism, 
as observed in experiments~\cite{Machida-2002,Moritomo-2004}.
Taking all these observation into consideration we classify different dopants 
in Table-1. 

\begin{table}[!t]
\centering
\begin{tabular}{|l|l|l|l|l|}
\hline
{}& {\bf $V and J'$} & {\bf Dopants} &{\bf {\it d} electrons}& {\bf \hspace{0.3cm} Type}\\ 
{}& {\bf } & {\bf (similar to)} &{\bf  }& {\bf }\\ \hline
{(i)}& $V=2, J'=0.15$ & Cr like    & \hspace{0.6cm} $d^{3}$ & Magnetic \\ \hline
{(ii)}& $V=1, J'=0.00$ & Ni like    & \hspace{0.6cm} $d^{7}$ & Magnetic  \\ \hline
{(iii)}& $V=5, J'=0.05$ & Fe like    & \hspace{0.6cm} $d^{5}$ & Magnetic   \\ \hline
{(iv)}& $V=5, J'=0.00$ & Ga/Al like & \hspace{0.6cm}  $d^{0}$ & Non-Magnetic\\ \hline
\end{tabular}
\caption{Various parameter $V$ and $J'$ to mimic different B-site 
dopants in experiments.}
\label{table-1}
\end{table}

To draw parallel with the experiments we show ferromagnetic structure
factor $S(\textbf{0,0})$ (thermal average combined with an additional 
average overten different `samples') with temperature for four combinations 
of $V$ and $J'$ values to probe different parts of the phase diagram in 
Fig-\ref{vj_fm_expt.eps}. Only first two combinations 
[$V=2, J'=0.15$ (Cr like); $V=1, J'=0.00$ (Ni like)] show significant 
ferromagnetism at low temperature while the other two have negligible 
ferromagnetic correlations. 

In conclusion, we explained the non-trivial effect of B-site dopants on a 
CE-CO-OO-I phase. Although valence change is in opposite direction with 
respect to the FM-M phase, the magnetic reconstructions create large 
magnetic patches and induce ferromagnetism for Cr/Ni like dopants. 
We also show that there are two types of magnetic reconstructions 
that induce ferromagnetism depending upon the magnetic interaction between 
the B-site dopants and the Mn-sites. We crudely guessed the value of SE 
interaction between the B-site dopants and neighboring Mn-sites, but we believe 
one can extract this value from first principle calculations. 

We acknowledge our discussions with Pinaki Majumdar and Anamitra Mukherjee.



\begin{thebibliography}{99}


\bibitem{Dagotto-2005} 
E. Dagotto, Science, {\bf 309}, 257 (2005).

\bibitem{Tokura-2000}
{\it {Colossal Magnetoresistive Oxides}}, edited by Y. Tokura (Gordon and
Breach, New York, 2000).

\bibitem{Tokura-2006} 
Y. Tokura,  Rep. Prog. Phys. {\bf 69}, 797 (2006).

\bibitem{Attfield-1998}
L. M. Rodriguez-Martinez and J. P. Attfield, 
Phys. Rev. {\bf B 58}, 2426 (1998).

\bibitem{Attfield-2001}
L. M. Rodriguez-Martinez and J. P. Attfield, 
Phys. Rev. {\bf B 63}, 024424 (2001).

\bibitem{Akahoshi-2003}
D. Akahoshi, M. Uchida, Y. Tomioka, T. Arima, Y. Matsui, and Y. Tokura, 
Phys. Rev. Lett. {\bf 90}, 177203 (2003).

\bibitem{Mathieu-2004} 
R. Mathieu, D. Akahoshi, A. Asamitsu, Y. Tomioka, and Y. Tokura, 
Phys. Rev. Lett. {\bf 93}, 227202 (2004).

\bibitem{Moreo-2000}
A. Moreo, M. Mayr, A. Feiguin, S. Yunoki, and 
E. Dagotto, Phys. Rev. Lett. {\bf 84}, 5568 (2000).

\bibitem{Kumar-2004} 
S. Kumar and P. Majumdar, Phys. Rev. Lett. {\bf 92}, 126602 (2004).

\bibitem{Kimura-1999}  
T. Kimura, Y. Tomioka, R. Kumai, Y. Okimoto, and Y. Tokura, 
Phys. Rev. Lett. {\bf 83}, 3940 (1999).

\bibitem{Kimura-2000}  
T. Kimura, R. Kumai, Y. Okimoto, and Y. Tomioka, 
Phys. Rev. B {\bf 62}, 15021 (2000).

\bibitem{Moritomo-1999} 
Y. Moritomo, A. Machida, S. Mori, N. Yamamoto, and A. Nakamura, 
Phys. Rev. B {\bf 60}, 9220 (1999).

\bibitem{Mori-2003} 
S. Mori,  R. Shoji, N. Yamamoto, T. Asaka, Y. Matsui, A. Machida, 
Y. Moritomo, and T. Katsufuji, Phys. Rev. {\bf B 67}, 012403 (2003).

\bibitem{Moritomo-2004} 
Y. Moritomo, K. Murakami, H. Ishikawa, M. Hanawa, A. Nakamura, 
and K. Ohoyama, Phys. Rev. B {\bf 69}, 212407 (2004).

\bibitem{Machida-2002} 
A. Machida, Y. Moritomo, K. Ohoyama, T. Katsufuji, and A. Nakamura, 
Phy. Rev. {\bf B 65}, 064435 (2002).

\bibitem{Ahn-1997}  
K. H. Ahn, X. W. Wu, K. Liu, and C. L. Chien, 
J. Appl. Phys. {\bf 81}, 5505 (1997).

\bibitem{Barnabe-1997} 
A. Barnabe, A. Maignan, M. Hervieu, F. Damay, C. Martin, and 
B. Raveau, Appl. Phys. Lett. {\bf 71}, 3907 (1997).

\bibitem{Sakai-2008}  
H. Sakai, K. Ito, T. Nishiyama, Xiuzhen Yu, Y. Matsui, S. Miyasaka, 
Y. Tokura, J. Phys. Soc. Jpn. {\bf 77}, 124712 (2008).

\bibitem{Dhiman-2013} 
Indu Dhiman, A. Das, A. K. Nigam, R. K. Kremer,
Journal of Magnetism and Magnetic Materials, {\bf 334}, 21 (2013).

\bibitem{Lu-2014} 
Chengliang Lu, Ni Hu, Ming Yang, S. Xia, H. Wang, J. Wang, Z. Xia, 
and J. M. Liub, Sci. Rep., {\bf 4}, 4902 (2014).

\bibitem{Pradhan-2008} 
K. Pradhan, A. Mukherjee and P. Majumdar,
Europhys. Lett. {\bf 84}, 37007 (2008).

\bibitem{Hebert-2002} 
S. Hebert, A. Maignan, C. Martin and B. Raveau, Solid State Commun. 
{\bf 121} 229 (2002).

\bibitem{Dagotto-2001} 
E. Dagotto, T. Hotta, and A. Moreo, 
Phys. Rep. {\bf 344}, 1 (2001).

\bibitem{Dagotto-1998}  
E. Dagotto, S. Yunoki, A. L. Malvezzi, A. Moreo, 
J. Hu, S. Capponi, D. Poilblanc, and N. Furukawa, 
Phys. Rev. {\bf B 58},6414 (1998). 

\bibitem{Pradhan-2007} K. Pradhan, A. Mukherjee and P. Majumdar,
Phys. Rev. Lett. {\bf 99}, 147206 (2007).

\bibitem{1-Kumar-2006}  
S. Kumar and P. Majumdar, Eur. Phys. J. {\bf B 50}, 571 (2006).

\bibitem{3-Kumar-2006}
Sanjeev Kumar, Arno P. Kampf, and Pinaki Majumdar, 
Phys. Rev. Lett. {\bf 97}, 176403 (2006).

\bibitem{Kumar-2008} 
Sanjeev Kumar and Arno P. Kampf, 
Phys. Rev. Lett. {\bf 100}, 076406 (2008).

\bibitem{Pradhan-2013} 
K. Pradhan, Arno P. Kampf, 
Phys. Rev. {\bf B 88}, 155136 (2013).

\bibitem{Yunoki-2000}
Seiji Yunoki, Takashi Hotta, and Elbio Dagotto, 
Phys. Rev. Lett. {\bf 84}, 3714 (2000).

\bibitem{Fratini-2001}
S. Fratini, D. Feinberg, and M. Grilli, 
Eur. Phys. J. {\bf B 22}, 205110 (2001).

\bibitem{Shenoy-2007}
Vijay B. Shenoy, Tribikram Gupta, H. R. Krishnamurthy, and 
T. V. Ramakrishnan, Phys. Rev. Lett. {\bf 98}, 097201 (2007).

\bibitem{Studer-1999} 
F. studer, O. Toulemonde, J. Goedkoop, A. Barnabe, and B. Raveau,
Jpn. J. Appl. Phys. {\bf 38}, 377 (1999).

\bibitem{SE-U}
The SE interaction is dependent on $t$ scale 
and Hubbard repulsion U as $t^2/U$. In case of Fe$^{3+}$ (t$^3_{2g}$e$^2_g$), 
the U is large as the next electron goes to a t$_{2g}$ level. 
For Cr$^{3+}$ (t$^3_{2g}$e$^0_g$) the effective U is probably smaller, 
which is why Cr has a stronger SE coupling to Mn 
than Fe does.
 
\end{thebibliography}
\end{document}